\documentclass[12pt,thmsa]{article}
\usepackage{amssymb}

\makeatletter
\newcount\@hour\newcount\@minute\chardef\@x10\chardef\@xv60
\def\tcitime{
\def\@time{%
  \@minute\time\@hour\@minute\divide\@hour\@xv
  \ifnum\@hour<\@x 0\fi\the\@hour:%
  \multiply\@hour\@xv\advance\@minute-\@hour
  \ifnum\@minute<\@x 0\fi\the\@minute
  }}%

\@ifundefined{hyperref}{}{}

\@ifundefined{qExtProgCall}{\def\qExtProgCall#1#2#3#4#5#6{\relax}}{}
\def\QCTOpt[#1]#2{%
  \def\QCTOptB{#1}
  \def\QCTOptA{#2}
}
\def\QCTNOpt#1{%
  \def\QCTOptA{#1}
  \let\QCTOptB\empty
}
\def\Qct{%
  \@ifnextchar[{%
    \QCTOpt}{\QCTNOpt}
}
\def\QCBOpt[#1]#2{%
  \def\QCBOptB{#1}
  \def\QCBOptA{#2}
}
\def\QCBNOpt#1{%
  \def\QCBOptA{#1}
  \let\QCBOptB\empty
}
\def\Qcb{%
  \@ifnextchar[{%
    \QCBOpt}{\QCBNOpt}
}
\def\PrepCapArgs{%
  \ifx\QCBOptA\empty
    \ifx\QCTOptA\empty
      {}%
    \else
      \ifx\QCTOptB\empty
        {\QCTOptA}%
      \else
        [\QCTOptB]{\QCTOptA}%
      \fi
    \fi
  \else
    \ifx\QCBOptA\empty
      {}%
    \else
      \ifx\QCBOptB\empty
        {\QCBOptA}%
      \else
        [\QCBOptB]{\QCBOptA}%
      \fi
    \fi
  \fi
}
\newcount\GRAPHICSTYPE
\GRAPHICSTYPE=\z@
\def\GRAPHICSPS#1{%
 \ifcase\GRAPHICSTYPE
   \special{ps: #1}%
 \or
   \special{language "PS", include "#1"}%
 \fi
}%
\def\graffile#1#2#3#4{%
    \leavevmode
    \raise -#4 \BOXTHEFRAME{%
        \hbox to #2{\raise #3\hbox to #2{\null #1\hfil}}}%
}%
\def\draftbox#1#2#3#4{%
 \leavevmode\raise -#4 \hbox{%
  \frame{\rlap{\protect\tiny #1}\hbox to #2%
   {\vrule height#3 width\z@ depth\z@\hfil}%
  }%
 }%
}%
\newcount\draft
\draft=\z@

\newif\ifwasdraft
\wasdraftfalse

\def\GRAPHIC#1#2#3#4#5{%
 \ifnum\draft=\@ne\draftbox{#2}{#3}{#4}{#5}%
  \else\graffile{#1}{#3}{#4}{#5}%
  \fi
 }%
\def\addtoLaTeXparams#1{%
    \edef\LaTeXparams{\LaTeXparams #1}}%
\newif\ifBoxFrame \BoxFramefalse
\newif\ifOverFrame \OverFramefalse
\newif\ifUnderFrame \UnderFramefalse

\def\BOXTHEFRAME#1{%
   \hbox{%
      \ifBoxFrame
         \frame{#1}%
      \else
         {#1}%
      \fi
   }%
}

\def\doFRAMEparams#1{\BoxFramefalse\OverFramefalse\UnderFramefalse\readFRAMEparams#1\end}%
\def\readFRAMEparams#1{%
 \ifx#1\end%
  \let\next=\relax
  \else
  \ifx#1i\dispkind=\z@\fi
  \ifx#1d\dispkind=\@ne\fi
  \ifx#1f\dispkind=\tw@\fi
  \ifx#1t\addtoLaTeXparams{t}\fi
  \ifx#1b\addtoLaTeXparams{b}\fi
  \ifx#1p\addtoLaTeXparams{p}\fi
  \ifx#1h\addtoLaTeXparams{h}\fi
  \ifx#1X\BoxFrametrue\fi
  \ifx#1O\OverFrametrue\fi
  \ifx#1U\UnderFrametrue\fi
  \ifx#1w
    \ifnum\draft=1\wasdrafttrue\else\wasdraftfalse\fi
    \draft=\@ne
  \fi
  \let\next=\readFRAMEparams
  \fi
 \next
 }%
\def\IFRAME#1#2#3#4#5#6{%
      \bgroup
      \let\QCTOptA\empty
      \let\QCTOptB\empty
      \let\QCBOptA\empty
      \let\QCBOptB\empty
      #6%
      \parindent=0pt%
      \leftskip=0pt
      \rightskip=0pt
      \setbox0 = \hbox{\QCBOptA}%
      \@tempdima = #1\relax
      \ifOverFrame
          \typeout{This is not implemented yet}%
          \show\HELP
      \else
         \ifdim\wd0>\@tempdima
            \advance\@tempdima by \@tempdima
            \ifdim\wd0 >\@tempdima
               \textwidth=\@tempdima
               \setbox1 =\vbox{%
                  \noindent\hbox to \@tempdima{\hfill\GRAPHIC{#5}{#4}{#1}{#2}{#3}\hfill}\\%
                  \noindent\hbox to \@tempdima{\parbox[b]{\@tempdima}{\QCBOptA}}%
               }%
               \wd1=\@tempdima
            \else
               \textwidth=\wd0
               \setbox1 =\vbox{%
                 \noindent\hbox to \wd0{\hfill\GRAPHIC{#5}{#4}{#1}{#2}{#3}\hfill}\\%
                 \noindent\hbox{\QCBOptA}%
               }%
               \wd1=\wd0
            \fi
         \else
            \ifdim\wd0>0pt
              \hsize=\@tempdima
              \setbox1 =\vbox{%
                \unskip\GRAPHIC{#5}{#4}{#1}{#2}{0pt}%
                \break
                \unskip\hbox to \@tempdima{\hfill \QCBOptA\hfill}%
              }%
              \wd1=\@tempdima
           \else
              \hsize=\@tempdima
              \setbox1 =\vbox{%
                \unskip\GRAPHIC{#5}{#4}{#1}{#2}{0pt}%
              }%
              \wd1=\@tempdima
           \fi
         \fi
         \@tempdimb=\ht1
         \advance\@tempdimb by \dp1
         \advance\@tempdimb by -#2%
         \advance\@tempdimb by #3%
         \leavevmode
         \raise -\@tempdimb \hbox{\box1}%
      \fi
      \egroup%
}%
\def\DFRAME#1#2#3#4#5{%
 \begin{center}
     \let\QCTOptA\empty
     \let\QCTOptB\empty
     \let\QCBOptA\empty
     \let\QCBOptB\empty
     \ifOverFrame 
        #5\QCTOptA\par
     \fi
     \GRAPHIC{#4}{#3}{#1}{#2}{\z@}
     \ifUnderFrame 
        \nobreak\par #5\QCBOptA
     \fi
 \end{center}%
 }%
\def\FFRAME#1#2#3#4#5#6#7{%
 \begin{figure}[#1]%
  \let\QCTOptA\empty
  \let\QCTOptB\empty
  \let\QCBOptA\empty
  \let\QCBOptB\empty
  \ifOverFrame
    #4
    \ifx\QCTOptA\empty
    \else
      \ifx\QCTOptB\empty
        \caption{\QCTOptA}%
      \else
        \caption[\QCTOptB]{\QCTOptA}%
      \fi
    \fi
    \ifUnderFrame\else
      \label{#5}%
    \fi
  \else
    \UnderFrametrue%
  \fi
  \begin{center}\GRAPHIC{#7}{#6}{#2}{#3}{\z@}\end{center}%
  \ifUnderFrame
    #4
    \ifx\QCBOptA\empty
      \caption{}%
    \else
      \ifx\QCBOptB\empty
        \caption{\QCBOptA}%
      \else
        \caption[\QCBOptB]{\QCBOptA}%
      \fi
    \fi
    \label{#5}%
  \fi
  \end{figure}%
 }%
\newcount\dispkind%

\def\makeactives{
  \catcode`\"=\active
  \catcode`\;=\active
  \catcode`\:=\active
  \catcode`\'=\active
  \catcode`\~=\active
}
\bgroup
   \makeactives
   \gdef\activesoff{%
      \def"{\string"}
      \def;{\string;}
      \def:{\string:}
      \def'{\string'}
      \def~{\string~}
    }
\egroup

\def\FRAME#1#2#3#4#5#6#7#8{%
 \bgroup
 \@ifundefined{bbl@deactivate}{}{\activesoff}
 \ifnum\draft=\@ne
   \wasdrafttrue
 \else
   \wasdraftfalse%
 \fi
 \def\LaTeXparams{}%
 \dispkind=\z@
 \def\LaTeXparams{}%
 \doFRAMEparams{#1}%
 \ifnum\dispkind=\z@\IFRAME{#2}{#3}{#4}{#7}{#8}{#5}\else
  \ifnum\dispkind=\@ne\DFRAME{#2}{#3}{#7}{#8}{#5}\else
   \ifnum\dispkind=\tw@
    \edef\@tempa{\noexpand\FFRAME{\LaTeXparams}}%
    \@tempa{#2}{#3}{#5}{#6}{#7}{#8}%
    \fi
   \fi
  \fi
  \ifwasdraft\draft=1\else\draft=0\fi{}%
  \egroup
 }%
\def\TEXUX#1{"texux"}

\long\def\QQQ#1#2{%
     \long\expandafter\def\csname#1\endcsname{#2}}%
\@ifundefined{QTP}{\def\QTP#1{}}{}
\@ifundefined{QEXCLUDE}{\def\QEXCLUDE#1{}}{}
\@ifundefined{Qlb}{}{}
\@ifundefined{Qlt}{}{}
\long\def\QQA#1#2{}%
\def\QTR#1#2{{\csname#1\endcsname #2}}
\def\EXPAND#1[#2]#3{}%
\def\NOEXPAND#1[#2]#3{}%
\def\LaTeXparent#1{}%
\def\ChildStyles#1{}%
\def\ChildDefaults#1{}%
\def\QTagDef#1#2#3{}%
\@ifundefined{StyleEditBeginDoc}{}{}
\def\QQfnmark#1{\footnotemark}

\@ifundefined{INDEX}{\def\INDEX#1#2{}{}}{}%
\@ifundefined{SUBINDEX}{\def\SUBINDEX#1#2#3{}{}{}}{}%
\@ifundefined{initial}%
   {\def\initial#1{\bigbreak{\raggedright\large\bf #1}\kern 2\p@\penalty3000}}%
   {}%
\@ifundefined{entry}{}{}%
\@ifundefined{primary}{}{}%
\@ifundefined{secondary}{}{}%
\@ifundefined{ZZZ}{}{\makeatletter\input gnuindex.sty\makeatother\makeindex\makeatletter}%
\@ifundefined{abstract}{%
 \def\abstract{%
  \if@twocolumn
   \section*{Abstract (Not appropriate in this style!)}%
   \else \small 
   \begin{center}{\bf Abstract\vspace{-.5em}\vspace{\z@}}\end{center}%
   \quotation 
   \fi
  }%
 }{%
 }%
\@ifundefined{endabstract}{\def\endabstract
  {\if@twocolumn\else\endquotation\fi}}{}%
\@ifundefined{maketitle}{\def\maketitle#1{}}{}%
\@ifundefined{affiliation}{\def\affiliation#1{}}{}%
\@ifundefined{proof}{}{}%
\@ifundefined{endproof}{}{}%
\@ifundefined{newfield}{\def\newfield#1#2{}}{}%
\@ifundefined{chapter}{\def\chapter#1{\par(Chapter head:)#1\par }%
 \newcount\c@chapter}{}%
\@ifundefined{part}{\def\part#1{\par(Part head:)#1\par }}{}%
\@ifundefined{section}{\def\section#1{\par(Section head:)#1\par }}{}%
\@ifundefined{subsection}{\def\subsection#1%
 {\par(Subsection head:)#1\par }}{}%
\@ifundefined{subsubsection}{\def\subsubsection#1%
 {\par(Subsubsection head:)#1\par }}{}%
\@ifundefined{paragraph}{\def\paragraph#1%
 {\par(Subsubsubsection head:)#1\par }}{}%
\@ifundefined{subparagraph}{\def\subparagraph#1%
 {\par(Subsubsubsubsection head:)#1\par }}{}%
\@ifundefined{therefore}{}{}%
\@ifundefined{backepsilon}{}{}%
\@ifundefined{yen}{}{}%
\@ifundefined{registered}{%
   \def\registered{\relax\ifmmode{}\r@gistered
                    \else$\m@th\r@gistered$\fi}%
 \def\r@gistered{^{\ooalign
  {\hfil\raise.07ex\hbox{$\scriptstyle\rm\text{R}$}\hfil\crcr
  \mathhexbox20D}}}}{}%
\@ifundefined{Eth}{}{}%
\@ifundefined{eth}{}{}%
\@ifundefined{Thorn}{}{}%
\@ifundefined{thorn}{}{}%
\def\TEXTsymbol#1{\mbox{$#1$}}%
\@ifundefined{degree}{}{}%
\newdimen\theight
\def\Column{%
 \vadjust{\setbox\z@=\hbox{\scriptsize\quad\quad tcol}%
  \theight=\ht\z@\advance\theight by \dp\z@\advance\theight by \lineskip
  \kern -\theight \vbox to \theight{%
   \rightline{\rlap{\box\z@}}%
   \vss
   }%
  }%
 }%
\def\qed{%
 \ifhmode\unskip\nobreak\fi\ifmmode\ifinner\else\hskip5\p@\fi\fi
 \hbox{\hskip5\p@\vrule width4\p@ height6\p@ depth1.5\p@\hskip\p@}%
 }%
\def\miss{\hbox{\vrule height2\p@ width 2\p@ depth\z@}}%
\def\tcol#1{{\baselineskip=6\p@ \vcenter{#1}} \Column}  %
\def\newfmtname{LaTeX2e}
\def\chkcompat{%
   \if@compatibility
   \else
     \usepackage{latexsym}
   \fi
}

\ifx\fmtname\newfmtname
  \DeclareOldFontCommand{\rm}{\normalfont\rmfamily}{\mathrm}
  \DeclareOldFontCommand{\sf}{\normalfont\sffamily}{\mathsf}
  \DeclareOldFontCommand{\tt}{\normalfont\ttfamily}{\mathtt}
  \DeclareOldFontCommand{\bf}{\normalfont\bfseries}{\mathbf}
  \DeclareOldFontCommand{\it}{\normalfont\itshape}{\mathit}
  \DeclareOldFontCommand{\sl}{\normalfont\slshape}{\@nomath\sl}
  \DeclareOldFontCommand{\sc}{\normalfont\scshape}{\@nomath\sc}
  \chkcompat
\fi

\def\alpha{{\Greekmath 010B}}%
\def\beta{{\Greekmath 010C}}%
\def\gamma{{\Greekmath 010D}}%
\def\delta{{\Greekmath 010E}}%
\def\epsilon{{\Greekmath 010F}}%
\def\zeta{{\Greekmath 0110}}%
\def\eta{{\Greekmath 0111}}%
\def\theta{{\Greekmath 0112}}%
\def\iota{{\Greekmath 0113}}%
\def\kappa{{\Greekmath 0114}}%
\def\lambda{{\Greekmath 0115}}%
\def\mu{{\Greekmath 0116}}%
\def\nu{{\Greekmath 0117}}%
\def\xi{{\Greekmath 0118}}%
\def\pi{{\Greekmath 0119}}%
\def\rho{{\Greekmath 011A}}%
\def\sigma{{\Greekmath 011B}}%
\def\tau{{\Greekmath 011C}}%
\def\upsilon{{\Greekmath 011D}}%
\def\phi{{\Greekmath 011E}}%
\def\chi{{\Greekmath 011F}}%
\def\psi{{\Greekmath 0120}}%
\def\omega{{\Greekmath 0121}}%
\def\varepsilon{{\Greekmath 0122}}%
\def\vartheta{{\Greekmath 0123}}%
\def\varpi{{\Greekmath 0124}}%
\def\varrho{{\Greekmath 0125}}%
\def\varsigma{{\Greekmath 0126}}%
\def\varphi{{\Greekmath 0127}}%

\def\nabla{{\Greekmath 0272}}
\def\FindBoldGroup{%
   {\setbox0=\hbox{$\mathbf{x\global\edef\theboldgroup{\the\mathgroup}}$}}%
}

\def\Greekmath#1#2#3#4{%
    \if@compatibility
        \ifnum\mathgroup=\symbold
           \mathchoice{\mbox{\boldmath$\displaystyle\mathchar"#1#2#3#4$}}%
                      {\mbox{\boldmath$\textstyle\mathchar"#1#2#3#4$}}%
                      {\mbox{\boldmath$\scriptstyle\mathchar"#1#2#3#4$}}%
                      {\mbox{\boldmath$\scriptscriptstyle\mathchar"#1#2#3#4$}}%
        \else
           \mathchar"#1#2#3#4%
        \fi 
    \else 
        \FindBoldGroup
        \ifnum\mathgroup=\theboldgroup 
           \mathchoice{\mbox{\boldmath$\displaystyle\mathchar"#1#2#3#4$}}%
                      {\mbox{\boldmath$\textstyle\mathchar"#1#2#3#4$}}%
                      {\mbox{\boldmath$\scriptstyle\mathchar"#1#2#3#4$}}%
                      {\mbox{\boldmath$\scriptscriptstyle\mathchar"#1#2#3#4$}}%
        \else
           \mathchar"#1#2#3#4%
        \fi     	    
	  \fi}

\newif\ifGreekBold  \GreekBoldfalse
\let\SAVEPBF=\pbf
\def\pbf{\GreekBoldtrue\SAVEPBF}%

\@ifundefined{theorem}{}{}
\@ifundefined{lemma}{}{}
\@ifundefined{corollary}{}{}
\@ifundefined{conjecture}{}{}
\@ifundefined{proposition}{}{}
\@ifundefined{axiom}{}{}
\@ifundefined{remark}{}{}
\@ifundefined{example}{}{}
\@ifundefined{exercise}{}{}
\@ifundefined{definition}{}{}

\@ifundefined{mathletters}{%
  \newcounter{equationnumber}  
  \def\mathletters{%
     \addtocounter{equation}{1}
     \edef\@currentlabel{\theequation}%
     \setcounter{equationnumber}{\c@equation}
     \setcounter{equation}{0}%
     \edef\theequation{\@currentlabel\noexpand\alph{equation}}%
  }
  
}{}

\@ifundefined{BibTeX}{%
    \def\BibTeX{{\rm B\kern-.05em{\sc i\kern-.025em b}\kern-.08em
                 T\kern-.1667em\lower.7ex\hbox{E}\kern-.125emX}}}{}%
\@ifundefined{AmS}%
    {\def\AmS{{\protect\usefont{OMS}{cmsy}{m}{n}%
                A\kern-.1667em\lower.5ex\hbox{M}\kern-.125emS}}}{}%
\@ifundefined{AmSTeX}{}{}%
\ifx\ds@amstex\relax
\makeatother\endinput
\else
   \@ifpackageloaded{amstex}%
      {\message{amstex already loaded}\makeatother\endinput}
      {}
   \@ifpackageloaded{amsgen}%
      {\message{amsgen already loaded}\makeatother\endinput}
      {}
\fi
\def\DN@{\def\next@}%
\def\eat@#1{}%
\let\DOTSI\relax
\def\RIfM@{\relax\ifmmode}%
\def\FN@{\futurelet\next}%
\newcount\intno@
\def\iint{\DOTSI\intno@\tw@\FN@\ints@}%
\def\iiint{\DOTSI\intno@\thr@@\FN@\ints@}%
\def\iiiint{\DOTSI\intno@4 \FN@\ints@}%
\def\idotsint{\DOTSI\intno@\z@\FN@\ints@}%
\def\ints@{\findlimits@\ints@@}%
\newif\iflimtoken@
\newif\iflimits@
\def\findlimits@{\limtoken@true\ifx\next\limits\limits@true
 \else\ifx\next\nolimits\limits@false\else
 \limtoken@false\ifx\ilimits@\nolimits\limits@false\else
 \ifinner\limits@false\else\limits@true\fi\fi\fi\fi}%
\def\multint@{\int\ifnum\intno@=\z@\intdots@                          
 \else\intkern@\fi                                                    
 \ifnum\intno@>\tw@\int\intkern@\fi                                   
 \ifnum\intno@>\thr@@\int\intkern@\fi                                 
 \int}
\def\multintlimits@{\intop\ifnum\intno@=\z@\intdots@\else\intkern@\fi
 \ifnum\intno@>\tw@\intop\intkern@\fi
 \ifnum\intno@>\thr@@\intop\intkern@\fi\intop}%
\def\intic@{%
    \mathchoice{\hskip.5em}{\hskip.4em}{\hskip.4em}{\hskip.4em}}%
\def\negintic@{\mathchoice
 {\hskip-.5em}{\hskip-.4em}{\hskip-.4em}{\hskip-.4em}}%
\def\ints@@{\iflimtoken@                                              
 \def\ints@@@{\iflimits@\negintic@
   \mathop{\intic@\multintlimits@}\limits                             
  \else\multint@\nolimits\fi                                          
  \eat@}
 \else                                                                
 \def\ints@@@{\iflimits@\negintic@
  \mathop{\intic@\multintlimits@}\limits\else
  \multint@\nolimits\fi}\fi\ints@@@}%
\def\intkern@{\mathchoice{\!\!\!}{\!\!}{\!\!}{\!\!}}%
\def\plaincdots@{\mathinner{\cdotp\cdotp\cdotp}}%
\def\intdots@{\mathchoice{\plaincdots@}%
 {{\cdotp}\mkern1.5mu{\cdotp}\mkern1.5mu{\cdotp}}%
 {{\cdotp}\mkern1mu{\cdotp}\mkern1mu{\cdotp}}%
 {{\cdotp}\mkern1mu{\cdotp}\mkern1mu{\cdotp}}}%
\def\RIfM@{\relax\protect\ifmmode}
\def\text{\RIfM@\expandafter\text@\else\expandafter\mbox\fi}
\let\nfss@text\text
\def\text@#1{\mathchoice
   {\textdef@\displaystyle\f@size{#1}}%
   {\textdef@\textstyle\tf@size{\firstchoice@false #1}}%
   {\textdef@\textstyle\sf@size{\firstchoice@false #1}}%
   {\textdef@\textstyle \ssf@size{\firstchoice@false #1}}%
   \glb@settings}

\def\textdef@#1#2#3{\hbox{{%
                    \everymath{#1}%
                    \let\f@size#2\selectfont
                    #3}}}
\newif\iffirstchoice@
\firstchoice@true
\def\Let@{\relax\iffalse{\fi\let\\=\cr\iffalse}\fi}%
\def\vspace@{\def\vspace##1{\crcr\noalign{\vskip##1\relax}}}%
\def\multilimits@{\bgroup\vspace@\Let@
 \baselineskip\fontdimen10 \scriptfont\tw@
 \advance\baselineskip\fontdimen12 \scriptfont\tw@
 \lineskip\thr@@\fontdimen8 \scriptfont\thr@@
 \lineskiplimit\lineskip
 \vbox\bgroup\ialign\bgroup\hfil$\m@th\scriptstyle{##}$\hfil\crcr}%
\def\Sb{_\multilimits@}%
\def\endSb{\crcr\egroup\egroup\egroup}%
\def\Sp{^\multilimits@}%

\newdimen\ex@
\def\rightarrowfill@#1{$#1\m@th\mathord-\mkern-6mu\cleaders
 \hbox{$#1\mkern-2mu\mathord-\mkern-2mu$}\hfill
 \mkern-6mu\mathord\rightarrow$}%
\def\leftarrowfill@#1{$#1\m@th\mathord\leftarrow\mkern-6mu\cleaders
 \hbox{$#1\mkern-2mu\mathord-\mkern-2mu$}\hfill\mkern-6mu\mathord-$}%
\def\leftrightarrowfill@#1{$#1\m@th\mathord\leftarrow
\mkern-6mu\cleaders
 \hbox{$#1\mkern-2mu\mathord-\mkern-2mu$}\hfill
 \mkern-6mu\mathord\rightarrow$}%
\def\overrightarrow{\mathpalette\overrightarrow@}%
\def\overrightarrow@#1#2{\vbox{\ialign{##\crcr\rightarrowfill@#1\crcr
 \noalign{\kern-\ex@\nointerlineskip}$\m@th\hfil#1#2\hfil$\crcr}}}%

\def\overleftarrow{\mathpalette\overleftarrow@}%
\def\overleftarrow@#1#2{\vbox{\ialign{##\crcr\leftarrowfill@#1\crcr
 \noalign{\kern-\ex@\nointerlineskip}$\m@th\hfil#1#2\hfil$\crcr}}}%
\def\overleftrightarrow{\mathpalette\overleftrightarrow@}%
\def\overleftrightarrow@#1#2{\vbox{\ialign{##\crcr
   \leftrightarrowfill@#1\crcr
 \noalign{\kern-\ex@\nointerlineskip}$\m@th\hfil#1#2\hfil$\crcr}}}%
\def\underrightarrow{\mathpalette\underrightarrow@}%
\def\underrightarrow@#1#2{\vtop{\ialign{##\crcr$\m@th\hfil#1#2\hfil
  $\crcr\noalign{\nointerlineskip}\rightarrowfill@#1\crcr}}}%

\def\underleftarrow{\mathpalette\underleftarrow@}%
\def\underleftarrow@#1#2{\vtop{\ialign{##\crcr$\m@th\hfil#1#2\hfil
  $\crcr\noalign{\nointerlineskip}\leftarrowfill@#1\crcr}}}%
\def\underleftrightarrow{\mathpalette\underleftrightarrow@}%
\def\underleftrightarrow@#1#2{\vtop{\ialign{##\crcr$\m@th
  \hfil#1#2\hfil$\crcr
 \noalign{\nointerlineskip}\leftrightarrowfill@#1\crcr}}}%

\def\qopnamewl@#1{\mathop{\operator@font#1}\nlimits@}
\let\nlimits@\displaylimits
\def\setboxz@h{\setbox\z@\hbox}

\def\varlim@#1#2{\mathop{\vtop{\ialign{##\crcr
 \hfil$#1\m@th\operator@font lim$\hfil\crcr
 \noalign{\nointerlineskip}#2#1\crcr
 \noalign{\nointerlineskip\kern-\ex@}\crcr}}}}

 \def\rightarrowfill@#1{\m@th\setboxz@h{$#1-$}\ht\z@\z@
  $#1\copy\z@\mkern-6mu\cleaders
  \hbox{$#1\mkern-2mu\box\z@\mkern-2mu$}\hfill
  \mkern-6mu\mathord\rightarrow$}
\def\leftarrowfill@#1{\m@th\setboxz@h{$#1-$}\ht\z@\z@
  $#1\mathord\leftarrow\mkern-6mu\cleaders
  \hbox{$#1\mkern-2mu\copy\z@\mkern-2mu$}\hfill
  \mkern-6mu\box\z@$}

\def\projlim{\qopnamewl@{proj\,lim}}
\def\injlim{\qopnamewl@{inj\,lim}}
\def\varinjlim{\mathpalette\varlim@\rightarrowfill@}
\def\varprojlim{\mathpalette\varlim@\leftarrowfill@}
\def\varliminf{\mathpalette\varliminf@{}}
\def\varliminf@#1{\mathop{\underline{\vrule\@depth.2\ex@\@width\z@
   \hbox{$#1\m@th\operator@font lim$}}}}
\def\varlimsup{\mathpalette\varlimsup@{}}
\def\varlimsup@#1{\mathop{\overline
  {\hbox{$#1\m@th\operator@font lim$}}}}

\def\stackunder#1#2{\mathrel{\mathop{#2}\limits_{#1}}}%
\begingroup \catcode `|=0 \catcode `[= 1
\catcode`]=2 \catcode `\{=12 \catcode `\}=12
\catcode`\\=12 
|gdef|@alignverbatim#1\end{align}[#1|end[align]]
|gdef|@salignverbatim#1\end{align*}[#1|end[align*]]

|gdef|@alignatverbatim#1\end{alignat}[#1|end[alignat]]
|gdef|@salignatverbatim#1\end{alignat*}[#1|end[alignat*]]

|gdef|@xalignatverbatim#1\end{xalignat}[#1|end[xalignat]]
|gdef|@sxalignatverbatim#1\end{xalignat*}[#1|end[xalignat*]]

|gdef|@gatherverbatim#1\end{gather}[#1|end[gather]]
|gdef|@sgatherverbatim#1\end{gather*}[#1|end[gather*]]

|gdef|@gatherverbatim#1\end{gather}[#1|end[gather]]
|gdef|@sgatherverbatim#1\end{gather*}[#1|end[gather*]]

|gdef|@multilineverbatim#1\end{multiline}[#1|end[multiline]]
|gdef|@smultilineverbatim#1\end{multiline*}[#1|end[multiline*]]

|gdef|@arraxverbatim#1\end{arrax}[#1|end[arrax]]
|gdef|@sarraxverbatim#1\end{arrax*}[#1|end[arrax*]]

|gdef|@tabulaxverbatim#1\end{tabulax}[#1|end[tabulax]]
|gdef|@stabulaxverbatim#1\end{tabulax*}[#1|end[tabulax*]]

|endgroup

\def\align{\@verbatim \frenchspacing\@vobeyspaces \@alignverbatim
You are using the "align" environment in a style in which it is not defined.}

\@namedef{align*}{\@verbatim\@salignverbatim
You are using the "align*" environment in a style in which it is not defined.}
\expandafter\let\csname endalign*\endcsname =\endtrivlist

\def\alignat{\@verbatim \frenchspacing\@vobeyspaces \@alignatverbatim
You are using the "alignat" environment in a style in which it is not defined.}

\@namedef{alignat*}{\@verbatim\@salignatverbatim
You are using the "alignat*" environment in a style in which it is not defined.}
\expandafter\let\csname endalignat*\endcsname =\endtrivlist

\def\xalignat{\@verbatim \frenchspacing\@vobeyspaces \@xalignatverbatim
You are using the "xalignat" environment in a style in which it is not defined.}

\@namedef{xalignat*}{\@verbatim\@sxalignatverbatim
You are using the "xalignat*" environment in a style in which it is not defined.}
\expandafter\let\csname endxalignat*\endcsname =\endtrivlist

\def\gather{\@verbatim \frenchspacing\@vobeyspaces \@gatherverbatim
You are using the "gather" environment in a style in which it is not defined.}

\@namedef{gather*}{\@verbatim\@sgatherverbatim
You are using the "gather*" environment in a style in which it is not defined.}
\expandafter\let\csname endgather*\endcsname =\endtrivlist

\def\multiline{\@verbatim \frenchspacing\@vobeyspaces \@multilineverbatim
You are using the "multiline" environment in a style in which it is not defined.}

\@namedef{multiline*}{\@verbatim\@smultilineverbatim
You are using the "multiline*" environment in a style in which it is not defined.}
\expandafter\let\csname endmultiline*\endcsname =\endtrivlist

\def\arrax{\@verbatim \frenchspacing\@vobeyspaces \@arraxverbatim
You are using a type of "array" construct that is only allowed in AmS-LaTeX.}

\def\tabulax{\@verbatim \frenchspacing\@vobeyspaces \@tabulaxverbatim
You are using a type of "tabular" construct that is only allowed in AmS-LaTeX.}

\@namedef{arrax*}{\@verbatim\@sarraxverbatim
You are using a type of "array*" construct that is only allowed in AmS-LaTeX.}
\expandafter\let\csname endarrax*\endcsname =\endtrivlist

\@namedef{tabulax*}{\@verbatim\@stabulaxverbatim
You are using a type of "tabular*" construct that is only allowed in AmS-LaTeX.}
\expandafter\let\csname endtabulax*\endcsname =\endtrivlist

\def\@@eqncr{\let\@tempa\relax
    \ifcase\@eqcnt \def\@tempa{& & &}\or \def\@tempa{& &}%
      \else \def\@tempa{&}\fi
     \@tempa
     \if@eqnsw
        \iftag@
           \@taggnum
        \else
           \@eqnnum\stepcounter{equation}%
        \fi
     \fi
     \global\tag@false
     \global\@eqnswtrue
     \global\@eqcnt\z@\cr}

 \def\endequation{%
     \ifmmode\ifinner 
      \iftag@
        \addtocounter{equation}{-1} 
        $\hfil
           \displaywidth\linewidth\@taggnum\egroup \endtrivlist
        \global\tag@false
        \global\@ignoretrue   
      \else
        $\hfil
           \displaywidth\linewidth\@eqnnum\egroup \endtrivlist
        \global\tag@false
        \global\@ignoretrue 
      \fi
     \else   
      \iftag@
        \addtocounter{equation}{-1} 
        \eqno \hbox{\@taggnum}
        \global\tag@false%
        $$\global\@ignoretrue
      \else
        \eqno \hbox{\@eqnnum}
        $$\global\@ignoretrue
      \fi
     \fi\fi
 } 

 \newif\iftag@ \tag@false
 
 \def\tag{\@ifnextchar*{\@tagstar}{\@tag}}
 \def\@tag#1{%
     \global\tag@true
     \global\def\@taggnum{(#1)}}
 \def\@tagstar*#1{%
     \global\tag@true
     \global\def\@taggnum{#1}%
}

\makeatother
\begin{document}

\title{Modular Localization and the Bootstrap-Formfactor Program}
\author{Bert Schroer \\
Institut f\"{u}r Theoretische Physik, FU-Berlin, Arnimallee 14, 14195 Berlin%
\\
e-mail : schroer@physik.fu-berlin.de \\
Temporary adress: CBPF, Rio de Janeiro, Brazil \\
e-mail:schroer@cat.cbpf.br}
\maketitle

\begin{abstract}
We extend the ''modular localization'' principle from free to interacting
theories and test its power for the special class of d=1+1 factorizing
models.
\end{abstract}

\tableofcontents

\section{{}Introduction}

Presently QFT presents itself as being formed of several parts which seem to
drift apart into different directions. On the one hand there is the standard
approach which is centered around renormalized perturbation theory and the
various quantization methods (canonical, functional). Enriched with
geometrical ideas it has led to recent (mainly) mathematical advances via
string theory and the Seiberg-Witten duality structure. On the other hand
there is the more algebra-based low dimensional approach which has led to
the construction of rich families of chiral conformal and integrable QFT.
The latter approach, although being somewhat conservative in its use of
physical principles, has nevertheless led to many startling results
concerning e.g. fusion of antiparticles from particles, confined objects and
solitons as being two sides of the same coin, and other extended (and
somewhat surprising) manifestations of the principle of ''nuclear
democracy''. Thirdly there is a very small group of theoreticians who find
it profitable to continue the structural investigations of algebraic QFT.

In fact the most interesting message of the low dimensional constructive
bootstrap-formfactor program seems to be that the emphasis on the scattering
matrix advocated way back by Heisenberg and later by Chew, Stapp and others,
was well founded. What went wrong in those early attempts was related to the
enforced and artificial separation from local QFT and the (cyclically
recurrent) ideologically motivated working hypothesis of a theory of
everything (in this case: everything minus gravity). The main theme of this
article is the realization that the S-matrix in algebraic QFT acquires a new
hitherto unknown pivotal role in the construction of local nets (whose
generators are local fields). It belongs to the foundation of the local
field theory (in its role as \textit{the} net invariant which carries local
modular information) as well as to its roof (in its role as describing
scattering observables), a truly vexing ''bootstrap'' situation. The fact
that in d=1+1 factorizable theories Chew's bootstrap ideas for the S-matrix
work without fields (but with the help of ''fusion'' and ''Yang-Baxter'') is
not due to the correctness of the underlying philosophy but rather to
undeserved luck: the physical rapidity scattering variable is at the same
time the uniformization parameter of the analytic properties \footnote{%
Even in d=1+1 the situation is very far removed from the desired uniqueness
of Chew's S-matrix approach.}. In higher dimensions or without the
factorization, Chew's program would fail without the use of local fields
(and it did fail). In that case an iterative procedure which corrects the
S-matrix together with a locality improvement of states and fields may have
a constructive chance, a situation which could be vaguely reminiscent of the
Hartree-Fock iteration in Schr\"{o}dinger theory..

In this note we propose a new concept \cite{S1} of ''modular localization''
which, as will be shown, is capable of reconquering the lost unity of QFT.
In particular, we will learn a new and very interesting lesson from the
d=1+1 formfactor program. Far from being a special ''exotic'' construction,
remote from any ''real'' QFT, this approach, if analyzed with general and
deep concepts related to the TCP theorem and the S-matrix (interpreted as an
invariant of a local net), reveals a surprising new and powerful
nonperturbative construction principle which, so we hope, may turn out to be
the basis of a future new iterative constructive approach. Locality of
observables and localization of states (always relative to the vacuum or
some other distinguished reference state) in QFT is a conditio sine qua non
for the physical interpretation (without any outside impositions). Global
topology as in the so called ''topological field (?) theories'' or the
vacuum structure assigned to certain effective potentials which does not
result from the local structure of real time QFT remain part of mathematics.

The fastest way to get a glimpse at the ''modular localization'' is to look
at the relation of the Wigner representation theory \cite{Wi} for positive
energy representations of the Poincar\'{e} group and free fields. Whereas in
Wigner's theory these irreducible representations in $H_{Wig}^{(m,s)}$are
uniquely specified by their mass and their spin (resp. helicity), this
uniqueness is lost if one passes to free fields in the Fock-space $%
H_{F}^{(m,s)}$. There are infinitely many free fields in Fock-space and they
constitute the linear part (in creation and annihilation operators) of a
huge local equivalence class of fields, the so called Borchers class $B(m,s)$
\cite{Haag}\cite{wightman} Any cyclic (with respect to the vacuum)
representative field from this class generates the same net of local von
Neumann algebras in $H_{F}$: 
\begin{equation}
\mathcal{O}\rightarrow \mathcal{A(O)}
\end{equation}
In fact the emerging picture of pointlike fields, that behave similar to
coordinates in differential geometry, was the prime motivation for
formulating algebraic QFT in terms of nets of algebras \cite{Haag}. In our
illustrative example\cite{S1} we regained the lost Wigner unicity on the
level of nets. For a detailed presentation of the physical motivations and
aims of this algebraic QFT we refer to a forthcoming article in Annals of
Physics \cite{S2}$.$

For the following it is important to understand the\textit{\ direct}
construction of this net in terms of the ''modular localization'' principle.
For that one uses the operators which Wigner's theory affiliates with a
reference wedge for which we take the x-t wedge: 
\begin{eqnarray}
W_{r} &:&\left| t\right| <x \\
\delta ^{it} &\equiv &U(\Lambda _{r}(2\pi t)),\quad j\equiv U(r_{r}) 
\nonumber
\end{eqnarray}
Here $\Lambda _{r}(\chi )$ and $r_{r}$ are the x-t Lorentz-boost and the x-t
reflection. The latter is represented by a antiunitary operator $j$ related
(via a $\pi $-rotation around the x-axis) to the TCP transformation $%
\vartheta .$ For a charged particle this requires doubling of the Wigner
space $H_{Wig}^{(m,s)}.$ The reflection commutes with the L-boost. It has
been shown elsewhere \cite{longo}\cite{S1} that the unbounded antilinear
involution: 
\begin{equation}
\mathit{s}=j\delta ^{\frac{1}{2}}
\end{equation}
from which $j$ and $\delta ^{\frac{1}{2}}$ can be recovered by polar
decomposition, can be used in order to define a real subspace $H_{R}:$%
\begin{equation}
H_{R}=\left\{ \psi \in H_{Wig}\mid \mathit{s}\psi =\psi \right\}
\end{equation}
The momentum space wave functions in $H_{R}$ are in the domain of $\delta ^{%
\frac{1}{2}}$ and hence have analytic properties in the rapidity variable $%
\theta $ associated to the standard wedge W$:$ $p_{r}=(\left| p\right|
cosh\theta ,\left| p\right| sinh\theta ,p_{2},p_{3}),$ $\left| p\right|
=(m^{2}+p_{2}^{2}+p_{3}^{2})^{\frac{1}{2}}$. For selfconjugate particles $%
H_{R}$ consists of analytic functions in the strip $0<\theta $ $<\pi $ for
which the two boundary values are related by a generalized reality
condition: 
\[
D^{(s)}(i\sigma _{2})\stackunder{\chi \rightarrow i\pi }{a.c.}\left[
D^{(s)}(R_{W}(\Lambda _{r}(\chi ),p_{r})\right] \psi (\theta +i\pi
,p_{2},p_{3})=\overline{\psi (\theta ,p_{2},p_{3})} 
\]
where $\psi $ is the 2s+1 component Wigner wave function, $R_{W}(\Lambda
_{r},p)$ the Wigner rotation associated with the boost $\Lambda $ on which
the analytic continuation $a.c.$ acts, and $D^{(s)}(i\sigma _{2})$ the
charge conjugation matrix. In the non-selfconjugate case (particles $\neq $
antiparticles) the reflection $j$ involves a flip in a doubled Wigner space. 
$H_{R}$ has a property which is called ''standard'' i.e.: 
\begin{equation}
H_{R}\cap iH_{R}=\left\{ 0\right\} ,\quad
H_{R}+iH_{R}\,\,dense\,\,in\,\,H_{Wig}
\end{equation}
and j transforms $H_{R}$ into its symplectic complement $H_{R}^{\prime }$
(the symplectic form is the imaginary part of the scalar product in $%
H_{Wig}) $ which in the case of integer spin representations (the
modifications for halfinteger spin are explained in \cite{S1}) is the same
as the geometric opposite wedge space: 
\begin{eqnarray}
H_{R}^{\prime } &\equiv &H_{R}(W_{r})^{\prime }=H_{R}(W_{r}^{\prime }) \\
W_{r}^{\prime } &\equiv &W_{r}^{opposite},\quad H_{R}(W)\cap H_{R}(W^{\prime
})=\left\{ 0\right\}  \nonumber
\end{eqnarray}
$\mathcal{P}$-covariance generates from $W_{r}$ a family of wedges $W=gW_{r}$
and associated $\mathit{s}(W),j(W),\delta (W)$ and $H_{R}(W)$ with the net
isotony property: 
\begin{equation}
H_{R}(W_{1})\subset H_{R}(W_{2})\quad W_{1}\subset W_{2}
\end{equation}
where the properness of the inclusion is a consequence of the positivity of
the energy (in fact equivalent to it, as it turns out) \cite{longo}. Wedge
localization in the Wigner theory (not to be confused with the Newton-Wigner
localization) \cite{S1} is the statement that the dense subspace (the
precise domain of $\mathit{s}$) $H_{R}+iH_{R}$ describes localization inside
the reference wedge. From QFT it is known that if one applies smeared local
fields $\phi $ with test functions which are supported in $W_{r}$ to the
vacuum, one obtains a dense set of $W$-localized state vectors. The relation
between the field algebras restricted to wedges and the Tomita-Takesaki
modular theory was first observed by Bisognano and Wichmann \cite{BiWi} and
later used by Sewell \cite{Sew} in order to obtain a structural
understanding of the Hawking-Unruh effect. What is new is that these spaces
allow (in the present case for free systems) for a very neat
characterization in terms of closed real subspaces whose position within the
total space contains the full information about localization regions.
Localization inside compact regions viz. double cones C (which are
inaccessible by direct geometric modular theory) may be defined in terms of
(dense if nontrivial) intersections: 
\begin{equation}
H_{R}(C)=\bigcap_{W\supset C}H_{R}(W)
\end{equation}
As already mentioned, the double cone localization is fulfilled in all
positive energy representations with halfinteger spin, but not in d=3+1 m=0
''continuous spin'' \cite{S1} representation and also not in massive d%
\TEXTsymbol{<}3+1 representations with ''anyonic''spin.

The important last step in the construction of a localized net of von
Neumann algebras is the application of the CCR (Weyl) and the CAR functor
which maps the net of real Hillbert subspaces into the net of algebras \cite
{longo}\cite{S1} 
\begin{eqnarray}
H_{Wig} &\rightarrow &B(H_{F})=alg\left\{ Weyl(f)\mid f\in \mathcal{S}%
(M_{4})\right\} \\
H_{R}(W) &\rightarrow &\mathcal{A}(W)=alg\left\{ Weyl(f)\mid f\in \mathcal{D}%
(W)\right\}  \nonumber \\
\check{S} &=&J\Delta ^{\frac{1}{2}},\quad J=e^{j},\quad \Delta =e^{\delta } 
\nonumber
\end{eqnarray}
These operators $\check{S},J$ and $\Delta ^{it}$ are the Tomita-Takesaki
operators of the T-T modular theory of the von Neumann algebra $\mathcal{A}%
(W)$ in ''standard position''. The inverse hat $\vee $ on S helps to
distinguish the Tomita involution from the later appearing scattering
S-operator, usually referred to as the S-matrix.

For the somewhat subtle point that the obstruction against the equality of
the geometric opposite with the modular opposite localization in the case of
halfinteger spin requires the introduction of a ''Klein twist'' and the CAR
functor we refer to the mentioned literature \cite{S1}.

\section{Modular Localization and Interaction}

In order to obtain a clue of how to incorporate an intrinsic notion of
interactions into this modular localization setting, we remind ourselves
that if we do use pointlike fields, the modular localization for free fields
agrees with what we get by applying the polynomial in the localization
region supported smeared fields. In contrast to the conventional
characterization of localization in terms of x-space pointlike fields, the
modular characterization solely works in the \textit{momentum-(Fock)space}
of the (incoming) \textit{free particles}. It attributes a physical
significance to the precise position of the Reeh-Schlieder \cite{Haag} dense
set of localized vectors and the change of this position resulting from the
change of localization region. In order to formulate the modular
localization principle in the case of interactions, one must take note of
the fact that the scattering matrix S of local QFT is the product of the
interacting TCP $\Theta $ with the free (incoming) TCP $\Theta _{0}$ and
(since the rotation by which the Tomita reflection $J$ differs from $\Theta $
is interaction-independent as all connected Poicar\'{e} transformations are
interaction-independent) we have: 
\begin{equation}
S=\Theta \cdot \Theta _{0},\quad S=J\cdot J_{0}
\end{equation}
and as a result we obtain for the Tomita involution: 
\begin{equation}
\check{S}=J\Delta ^{\frac{1}{2}}=SJ_{0}\Delta ^{\frac{1}{2}}=S\check{S}_{0}
\end{equation}
Again we may use covariance in order to obtain $\check{S}(W)$ and the
localization domain of $\check{S}(W)$ as $\mathcal{D}(\check{S}%
(W))=H_{R}(W)+iH_{R}(W)$ i.e. in terms of a net of closed real subspaces $%
H_{R}(W)$ of the incoming Fock space. However now the construction of an
associated von Neumann algebra is not clear since an ''interacting'' functor
from subspaces of the Fock space to von Neumann algebras is not known. We
will make some remarks (still short of a solution of this important problem)
in the concluding section and continue here with some more helpful comments
on modular localization of interacting state vectors.

As in the free case, the modular wedge localization does not use full
Einstein causality but only the so-called ''weak locality'', which is just a
reformulation of the TCP invariance \cite{wightman} Weakly local fields form
an equivalence class which is much bigger than the local Borchers class but
they are still associated to the same S-matrix (or rather the same TCP
operator). Actually the S in local quantum physics has two different
interpretations: S in its role to provide modular localization in
interacting theories, and S with the standard scattering interpretation in
terms of (nonlocal!) large time limits. There is \textit{no parallel} 
\textit{outside local quantum physics} to this state of affairs. Whereas all
concepts and properties which have been used hitherto in standard QFT
(perturbation theory, canonical formalism and path integrals) as e.g. time
ordering \footnote{%
There is a conspicuous absence of the time-ordering operation in the
bootstrap construction of factorizable field models. Instead the basic
objects are generalized formfactors i.e. sesquilinear forms on a dense set
of state vectors.} and interaction picture formalism, are shared by
nonrelativistic theories, modular localization is a new structural element
in local quantum physics \footnote{%
This characteristic modular structure lifts local quantum physics to a new
realm by its own which cannot be obtained by specialization from general
quantum theory.} and a characteristic property for Einstein causal quantum
physics. The simplest kind of interacting theories are those in which the
particle number is at least asymptotically (''on shell'') conserved i.e. $%
\left[ \check{S},\mathbf{N}_{in}\right] =0.$

In the next section we will briefly review the d=1+1 bootstrap-formfactor
program in a manner which facilitates the later application of modular
localization.

\section{The Bootstrap-Formfactor Program}

In this section we will meet a constructive approach for ''integrable''
d=1+1 QFT. Our first task is to obtain an intrinsic QFT understanding of
integrability in a way which avoids classical notions as e.g. complete sets
of conservation laws etc. For this purpose we note an important d=1+1
peculiarity.. Our generic expectation is that large spatial separation of
the center of wave packet of two particles in the elastic two-particle
scattering matrix leads to the weakening of scattering, or in momentum
space: 
\begin{equation}
\left\langle p_{1}^{^{\prime }}p_{2}^{^{\prime }}\left| S\right|
p_{1}p_{2}\right\rangle =\left\langle p_{1}^{^{\prime }}p_{2}^{^{\prime }}%
\mathbf{\mid }p_{1}p_{2}\right\rangle +\delta (p_{1}+p_{2}-p_{1}^{^{\prime
}}-p_{2}^{^{\prime }})T(p_{1}p_{2}p_{1}^{^{\prime }}p_{2}^{^{\prime }})
\end{equation}
where the identity contribution is more singular (has more $\delta $%
-factors) than the T-term and therefore the second term drops out in x-space
clustering. This argument fails precisely in d=1+1 and therefore the cluster
property of the S-matrix is not suitable in order to obtain an intrinsic
understanding of interaction. The two-particle S-matrix looses its higher
particle threshold structure, but it remains nontrivial (in distinction to
d=3+1). However for all higher particle scattering processes the behavior
for d=1+1 is qualitatively the same as in higher dimensions: the decreasing
threshold singularities (which decrease with increasing particle number) are
responsible for the spatial decrease. Therefore any d=1+1 QFT is expected to
have a limiting S$_{\lim }$-matrix which is purely elastic and solely
determined by the elastic two-particle $S^{(2)}$-matrix. The Yang-Baxter
relation results as a consistency relation for the elastic 3$\rightarrow $3
particle $S_{\lim }^{(3)}$-matrix. If this limiting S-matrix would again
correspond to a localizable QFT, we would have a new class division of QFT,
this time based on a long distance limit (which in some sense is opposite to
the scale invariant short distance limit). It is this \textit{(long
distance) class property \footnote{%
Although I know of no article in which this has been spelled out, its
pervasive presence behind the scene is is recognizable in some publications.}
which makes these factorizing models so fascinating}, as much as the
fascination of chiral conformal QFT results from their role of representing
short distance universality classes. In d=3+1 S$_{\lim }=1$ and therefore
the limiting theory is expected to maintain the same superselection rules
but in the ''interaction freeest'' possible way (literally free theories as
we will argue later on). Hence in d=1+1 we are invited to speculate on the
validity of the following commutative diagram: 
\begin{equation}
\mathcal{F}\,_{\searrow }^{\nearrow } 
\begin{array}{l}
\mathcal{F}_{ld} \\ 
\stackrel{\downarrow }{\mathcal{F}}_{sd}
\end{array}
\end{equation}
Here $ld(sd)$ labels the long (short) distance limits. There are also
arguments \cite{Zam} that with the help of a perturbative idea one may
ascend from $\mathcal{F}_{sd}$ to $\mathcal{F}_{ld}.$ It is however presenly
not clear how one can use the known properties of the $ld$ theories (i.e.
integrable models) in order to formulate a constructive program for the
nonintegrable members of the $ld$ equivalence class. We hope that our
modular localization principle (which is not restricted to factorizable
models) may turn out to be helpful for this purpose.

The constructive approach based on the bootstrap idea proceeds in two steps.
One first classifies unitary, crossing symmetric solutions of the
Yang-Baxter equations which fulfill certain minimal (or maximal, depending
on the viewpoint) requirements. Afterwards we use these factorizing
S-matrices together with the Watson equations (a notion from scattering
theory relating formfactors with the S-matrix) and analytic properties for
formfactors in order to compute the latter. One obtains the complete set of
multi-particle matrix elements of ''would be'' local fields, i.e. one
constructs the fields as sesquilinear forms. It is characteristic of this
method that one does not use the ''axiomatic'' properties of the beginning
of this section but rather less rigorously known momentum-space analytic
properties which, although certainly related to causality and spectral
properties, are more part of the LSZ+dispersion theoretic folklore than of
rigorous QFT. As long as one demonstrates at the end that the so obtained
fields fulfill local commutativity this is a legitimate procedure. It leaves
open the question whether there exists a more direct conceptual link between
the S-matrix and the local fields or rather the field independent local
nets. That this is indeed the case will be shown after the presentation of
the formfactor program.

\subsection{Properties of Factorizing S-Matrices}

Consider first the analytic structure of an elastic S-matrix for a scalar
neutral particle. In terms of the rapidity variable $\theta $: 
\begin{equation}
\left| p_{1,}p_{2}\right\rangle ^{out}=S\left| p_{1},p_{2}\right\rangle
^{in}=S_{el}(p_{1},p_{2})\left| p_{1},p_{2}\right\rangle ^{in}
\end{equation}
\begin{eqnarray}
S_{el}(p_{1},p_{2}) &=&:S(\theta ),\quad p_{i}=m(cosh\theta _{i},sinh\theta
_{i}),\quad \theta :=\left| \theta _{1}-\theta _{2}\right| \\
^{in}\left\langle p_{1}^{\prime },p_{2}^{\prime }\left| S\right|
p_{1},p_{2}\right\rangle ^{in} &=&S(\theta )^{in}\left\langle p_{1}^{\prime
},p_{2}^{\prime }\mid p_{1},p_{2}\right\rangle ^{in}  \nonumber
\end{eqnarray}
Usually the elastic S-matrix is written in terms of the invariant energy $%
s=(p_{1}+p_{2})^{2}=2m^{2}(1+cosh\theta )$ and the momentum transfer (not
independent in d=1+1) $t=(p_{1}-p_{2})^{2}=$.$2m^{2}(1-cosh\theta ).$ As a
result of undeserved fortune, the rapidity $\theta $ turns out to be a
uniformization variable for the real analytic S i.e. the complex s-plane
with the elastic cut in $s\geq 4m^{2}$ is dumped into the strip $0\leq
Im\theta \leq \pi $ and the S-matrix becomes a meromorphic function $%
S(\theta )$ with $S(-\theta )=S^{*}(\theta )=S^{-1}(\theta ).$ (unitarity).
Hence the strip $-\pi \leq \theta \leq \pi $ is the physical strip for $%
S(\theta )$. Crossing symmetry in our special (neutral) case means a
symmetry on the boundary of the strip: $\theta \rightarrow i\pi -\theta $.
Note that the presence of inelastic thresholds would destroy the
uniformization.

The factorization implies the operator relation: 
\begin{eqnarray}
&&S_{12}(p_{1},p_{2})S_{13}(p_{1},p_{3})S_{23}(p_{2},p_{3}) \\
&=&S_{23}(p_{2},p_{3})S_{13}(p_{1},p_{3})S_{12}(p_{1},p_{2})  \nonumber
\end{eqnarray}
According to Liouville's theorem, the only minimal solution (minimal number
of poles,smallest increase at $\infty $) for this scalar diagonal case is $%
S=\pm 1.$ More general solutions are obtained by placing bound-state poles
into the minimal solution. In order to maintain unitarity, the pole factor
must be of the form: 
\begin{equation}
P(\theta )=\frac{sinh\theta +isin\lambda }{sinh\theta -isin\lambda }
\end{equation}
Transforming back this pole at $\theta =i\lambda $ into the original
individual particle variables, we obtain the following parametrization in
terms of a center of mass and relative rapidity: 
\begin{eqnarray}
p_{1} &=&m\left( cosh(\chi +\frac{i\lambda }{2}),sinh(\chi +\frac{i\lambda }{%
2})\right) \\
p_{2} &=&m\left( cosh(\chi -\frac{i\lambda }{2}),sinh(\chi -\frac{i\lambda }{%
2})\right)  \nonumber
\end{eqnarray}
Clearly the two-particle bound state has the momentum: 
\begin{eqnarray}
p_{1,2} &=&\left( p_{1}+p_{2}\right) _{at\,bd.state}=2mcos\frac{\lambda }{2}%
\left( cosh\chi ,sinh\chi \right) \\
p_{1,2}^{2} &=&m_{2}^{2},\quad m_{2}=\frac{m}{2sin\frac{\lambda }{2}}%
sin\lambda  \nonumber
\end{eqnarray}
The ''fusion'' of particles may be extended. For a 3-particle bound state we
would look at the 3-particle S-matrix which, as a result of factorization
has the form: 
\begin{equation}
S^{(3)}(p_{1},p_{2},p_{3})=S(\theta _{12})S(\theta _{13})S(\theta _{23})
\end{equation}
We first fuse 1with 2 and simultaneously 2 with 3 as before. The center of
mass + relative rapidity parametrization yields: 
\begin{eqnarray}
p_{1} &=&m\left( cosh(\chi +i\lambda ),sinh(\chi +i\lambda )\right) \\
p_{2} &=&m\left( cosh\chi ,sinh\chi \right)  \nonumber \\
p_{3} &=&m\left( cosh(\chi -i\lambda ),sinh(\chi -i\lambda )\right) 
\nonumber
\end{eqnarray}
Again we get the mass of the 3-particle bound state by adding the zero
components in the $\chi =0$ frame: 
\begin{equation}
m_{3}=(p_{1}+p_{2}+p_{3})_{0}=m_{2}cos\frac{\lambda }{2}+mcos\lambda =2\frac{%
m}{2sin\frac{\lambda }{2}}sin\frac{3\lambda }{2}
\end{equation}
Induction then gives the general fusion mass formula: 
\begin{equation}
m_{n}=2\mu sin\frac{n\lambda }{2},\quad \mu =\frac{m}{2sin\frac{\lambda }{2}}
\end{equation}
We will meet such trigonometric fusion formulas later in algebraic QFT where
they are related to the statistical dimensions of fused charge sectors. They
were first known through the Dashen-Hasslacher-Neveu quasiclassical
approach. The above fusion calculation was done as far back as 1976 \cite
{STW} and consisted in a synthesis of the quasiclassical work of DHN with
some ideas of Sushko using the factorization principle, but still without
the ideas of Yang and Baxter (which are not needed for this scalar case).
The decisive step towards a general factorizable bootstrap program was taken
two years later \cite{Ka}\cite{Za}$.$

The consistency of these particles as incoming and outgoing objects leads to
additional structures. Consider the scattering of the mass $m_{2}$ bound
state with a third m-particle. This S-matrix for the scattering of these two
different particles is obtained from $S^{(3)}$ by: 
\begin{equation}
S_{b.e.}(p_{1}+p_{2},p_{3})\mid _{(p_{1}+p_{2})^{2}=m_{2}^{2}}=\frac{1}{R}%
\stackunder{(p_{1}+p_{2})^{2}\rightarrow m_{2}^{2}}{Res}S_{12}S_{13}S_{23}
\end{equation}
where the projector $P_{12}$ together with a numerical residue value $R$ is
defined by: 
\begin{equation}
\stackunder{(p_{1}+p_{2})^{2}\rightarrow m_{2}^{2}}{Res}%
S(p_{1},p_{2})=RP_{12}
\end{equation}
and we used the word elementary $e$.and bound $b.$ as labels on the new
two-particle $S_{b.c.}.$ The factorization insures that: 
\begin{equation}
P_{12}S_{13}S_{23}=S_{23}S_{13}P_{12}
\end{equation}
A prominent family of scalar S-matrices with N -1 bound state fulfilling all
these requirements are the $Z_{N}$ models \cite{Swieca}. Consistency
requires that the bound state of N-1 m-particles is again a m-particle. For
N=2 this family contains the Ising field theory with $S_{Ising}^{(2)}=-1$
which we already met in the section on (dis)order variables. Instead of
elaborating this scalar factorization situation, we pass immediatly to the
matrix case where we meet a new and interesting phenomenon. We assume that
the particle from which we start has an internal ''charge'' which can take
on a finite number of values i.e. 
\begin{equation}
\left| p,\alpha \right\rangle \in H_{1}\otimes V,\quad dimV<\infty
\end{equation}
The two-particle S-matrix is then written as a matrix acting on $V\otimes V$
whose entries are operator-valued (represented as in the previous case by
momentum-space kernels): 
\begin{equation}
S\left| p_{1},....,p_{n}\right\rangle _{\alpha _{1}....\alpha
_{n}}^{in}=\left| p_{1},....,p_{n}\right\rangle _{\alpha _{1}^{\prime
}....\alpha _{n}^{\prime }}^{in}S_{\alpha _{1}....\alpha _{n}}^{\alpha
_{1}^{\prime }....\alpha _{n}^{\prime }}(p_{1},....p_{n})
\end{equation}
\begin{equation}
S^{(n)}(p_{1},....,p_{n})=\prod_{i<j}S^{(2)}(p_{i},p_{j})
\end{equation}
The factorization requires a specific order of the product of matrices.
Consistency requires the validity of a Artin (braid-group) like relation: 
\begin{eqnarray}
&&S_{12}(p_{1},p_{2})S_{13}(p_{1},p_{3})S_{23}(p_{2},p_{3}) \\
&=&S_{23}(p_{2},p_{3})S_{13}(p_{1},p_{3})S_{12}(p_{1},p_{2})  \nonumber
\end{eqnarray}
The notation should be obvious: the subscript on S indicates on which of the
tensor factors in the 3-fold tensor product of one-particle spaces the
object acts. The relation with the Artin relations becomes clear if one
ignores the p-dependence and rewrites the Y-B relation in terms of $\tilde{S}%
.=PS$, where P is the permutation of two tensor factors.

This is the famous Yang-Baxter relation, since at the time of the discovery
of the S-matrix bootstrap it became clear that such a mathematical structure
had appeared before outside QFT in a quite different setting. Here this
identity permits to change the temporal order of individual rescatterings so
that the n-particle scattering $S^{(n)}$ is independent of those
(graphically: invariance under parallel shifts of 2-momenta in graphical
illustrations of scattering processes). The problem of finding the natural
parametrization (e.g. Baxter's elliptic parametrization) for these
Yang-Baxter relations does not arise in QFT; the \textit{uniformizing
rapidity }$\theta $\textit{\ is already the natural Yang-Baxter variable: } 
\begin{equation}
S_{12}(\theta )S_{13}(\theta +\theta ^{\prime })S_{23}(\theta ^{\prime
})=S_{23}(\theta ^{\prime })S_{13}(\theta +\theta ^{\prime })S_{12}(\theta )
\end{equation}
If fermion-antifermion pairs can go into boson-antiboson pairs, the object
which fulfills the Yang-Baxter relation is not S but $\sigma S$ where $%
\sigma =\pm 1$ with + for bosons. As the braid group relation, this is an
overdetermined system of equations. For the former one found a powerful
mathematical framework within V.Jones subfactor theory \cite{Jones}.
Although the attempts to get an equally powerful mathematical framework for
the latter was less than successful (the ''Baxterization'' of the subfactor
representations of Artin braids) one was able to find many interesting
families of nontrivial solutions of which some even allowed a comparision
with Lagrangian perturbation theory.

The S-matrix bootstrap idea originated in the early 60$^{ies}$ from
dispersion theory. Its revival in connection with d=1+1 factorization in the
late 70$^{ies}$ showed that its premises were physically reasonable, except
the idea that it could be seen as a ''theory of everything'' (TOE) which was
wrong and even obsert (for the more recent TOE's one would be hard pressed
to say friendly words about their physical content).

The basic new message \cite{Smir}\cite{BFK} is that one should use these
factorizing S-matrices as computational tools for the construction of local
fields and local nets as explained in the following subsection

\subsection{Generalized Formfactors}

Now we will probe the idea that these S-matrices belong to localizable
fields. Let A be any local field which belongs to a Borchers equivalence
class of local fields. We write the generalized formfactor of A(x) as: 
\begin{equation}
_{\alpha _{1}....\alpha _{m}}{}^{out}\left\langle p_{1},....,p_{m}\left|
A(0)\right| p_{m+1},....,p_{n}\right\rangle _{\alpha _{m+1}....\alpha
_{n}}^{in}  \label{mixed}
\end{equation}

We are interested in its analytic p-space properties. ''On shell'' p-space
analytic properties are more elusive than x-space analytic properties. For
the latter the spectral support properties play the important role, whereas
p-space analyticity relies heavily on caucality. The above matrix element
still contains energy-momentum $\delta $-functions resulting from
contracting incoming p's with outgoing. These are removed by taking the
connected parts of the formfactors. Only for the distinguished formfactor: 
\begin{equation}
\left\langle 0\left| A(0)\right| p_{1},....p_{n}\right\rangle _{\alpha
_{1}....\alpha _{n}}^{in}=\left\langle 0\left| A(0)\right|
p_{1},....p_{n}\right\rangle _{\alpha _{1}....\alpha _{n}}^{in,con}
\end{equation}
we have coalescence with its connected part. Similar to x-space analyticity,
one expects the existence of one analytic master-function whose different
boundary values correspond to the different n-particle formfactors: 
\begin{eqnarray}
&&_{\alpha _{1}....\alpha _{m}}{}^{out}\left\langle p_{1},....,p_{m}\left|
A(0)\right| p_{m+1},....,p_{n}\right\rangle _{\alpha _{m+1}....\alpha
_{n}}^{in,con} \\
&=&F_{\underline{\alpha }}^{A}(s_{ij}+i\varepsilon ,t_{rs}-i\varepsilon
,s_{kl}+i\varepsilon ),\quad i<j\leq m<k<l\leq n  \nonumber \\
t_{rs} &=&\left( p_{r}-p_{s}\right) ^{2},\quad r\leq m<s\leq n  \nonumber
\end{eqnarray}
There are Watson relations between the S-matrix and the formfactors. In the
d=1+3 dispersion theory setting it is well known that the cuts below the
inelastic threshold of $\left\langle 0\left| A(0)\right|
p_{1},p_{2}\right\rangle $ is related to the partial wave phase shifts in
that elastic region. In a factorizing d=1+1 theory these Watson relations
can be written down in general: 
\begin{eqnarray}
F_{\alpha _{1}....\alpha _{n}}^{A}(s_{ij}+i\varepsilon ) &=&\left\langle
0\left| A(0)\right| p_{1},....p_{n}\right\rangle _{\alpha _{1}....\alpha
_{n}}^{in} \\
&=&\sum_{out}\left\langle 0\left| A(0)\right| out\right\rangle \left\langle
out\mid p_{1},....,p_{n}\right\rangle _{\alpha _{1}....\alpha _{n}}^{in} 
\nonumber
\end{eqnarray}
\begin{equation}
\curvearrowright F_{\alpha _{1}....\alpha _{n}}^{A}(s_{ij}+i\varepsilon
)=F_{\alpha _{1}^{\prime }....\alpha _{n}^{\prime }}^{A}(s_{ij}-i\varepsilon
)S_{\alpha _{1}....\alpha _{n}}^{\alpha _{1}^{\prime }....\alpha
_{n}^{\prime }}(s_{ij})
\end{equation}
and for the mixed formfactors\ref{mixed}: 
\begin{eqnarray}
&&F_{\underline{\alpha }}^{A}(s_{ij}+i\varepsilon ,t_{rs}-i\varepsilon
,s_{kl}+i\varepsilon ) \\
&=&S_{\alpha _{1}^{\prime }....\alpha _{m}^{\prime }}^{\alpha _{m}....\alpha
_{1}}(s_{ij})F_{\underline{\alpha }^{\prime }}^{A}(s_{ij}-i\varepsilon
,t_{rs}+i\varepsilon ,s_{kl}-i\varepsilon )S_{\alpha _{m+1}....\alpha
_{n}}^{\alpha _{n}^{\prime }....\alpha _{m+1}^{\prime }}(s_{kl})  \nonumber
\end{eqnarray}
Using the uniformazing $\theta ^{\prime }s$ this is like a generalized
quasiperiodicity property on $\theta $-strips for the F's (instead of the
periodicity of S). The first who considered formfactors beyond two-particles 
\cite{KW} and presented a system of axioms for their calculation was Smirnov 
\cite{Smir} Following a recent presentation by Babujian, Fring and Karowski 
\cite{BFK} in a more standard field theoretic setting (LSZ+ dispersion
theory), the formfactor program for the construction of d=1+1 QFT is as
follows. Introduce the orderd formfactors: 
\begin{equation}
f_{\underline{\alpha }}^{A}(\theta _{1},....,\theta _{n}):=\left\langle
0\left| A(0)\right| p_{1},....,p_{n}\right\rangle _{\underline{\alpha }%
}^{in},\quad \theta _{1}>...>\theta _{n}  \label{def}
\end{equation}
and define the value for reordered $\theta ^{\prime }s$ by analytic
continuation (starting with this ordering in the physical region). Demand
that the f's fulfill the following properties:

\begin{itemize}
\item  (i)$\quad f_{...ij...}^{A}(...,\theta _{i},\theta
_{j},...)=f_{...ji...}^{A}(...,\theta _{j},\theta _{i},...)S_{ij}(\theta
_{i}-\theta _{j})\quad \forall \,\,\theta ^{\prime }s$

\item  (ii)\quad $f_{12...n}^{A}(\theta _{1}+i\pi ,\theta _{2},...,\theta
_{n})=f_{2...n1}^{A}(\theta _{2},...,\theta _{n},\theta _{1}-i\pi )$

\item  (iii)\quad $f_{1...n}^{A}(\theta _{1},...\theta _{n})\stackunder{%
\theta _{1}\rightarrow \theta _{2}+i\pi }{\approx }\frac{2i}{\theta
_{1}-\theta _{2}-i\pi }C_{12}f_{3...n}^{A}(\theta _{3},...,\theta
_{n})(1-S_{2n}...S_{23})$
\end{itemize}

where $C_{\alpha \beta }=\delta _{\bar{\alpha}\beta }$ is the charge
conjugation matrix.

Here we have not mentioned the poles from bound states (states which appear
by the previous fusion) since they are automatically entering the
formfactors via the S-matrix. The word ''axiom'' in the context of this
paper has the significance of working hypothesis i.e. an assumption which
receives its legitimation through its constructive success. Physical
principles on the other hand, as the spectral and causality properties of
general QFT, will not be called axioms. Our main aim is to show how one can
reduce the above axioms of the bootstrap-formfactor approach to the
principles of QFT and thereby recuperate the unity of this nonperturbative
approach with the rest of QFT.

The conceptually somewhat unusual property is the ''symmetry'' property (i).
Here one should bear in mind that from the point of view of the LSZ
formulation f is an auxiliary object to which the statistics property under
particle exchange does not apply (it would apply to the original
matrix-element). The above exchange property for $f$ is a statement about
analytic continuation. The statistics of incoming particle is only used in
order to get the charges (i.e. the tensor factors) into the same $j-i$ order
as the analytically interchanged $\theta ^{\prime }s.$ Following BFK \cite
{BFK} let us first remind ourselves of the standard argument for (i) in
somewhat detail. For the special case $\left\langle 0\left| A(0)\right|
\theta _{1}\theta _{2}\right\rangle ^{ex}$ $ex=in,out$ it is evident that: 
\begin{equation}
lim_{\varepsilon \rightarrow 0}F(s_{12}\pm i\varepsilon )=\left\{ 
\begin{array}{l}
\left\langle 0\left| A(0)\right| \theta _{1}\theta _{2}\right\rangle ^{in}
\\ 
\left\langle 0\left| A(0)\right| \theta _{1}\theta _{2}\right\rangle ^{out}
\end{array}
\right.
\end{equation}
i.e. there is one analytic masterfunction $f(z)$ (assuming identical
particles) with different boundary values on the $s\geq 4m^{2}$ cut having
the $in,out$ interpretation. Assuming Bose statistics, the physical matrix
elements on the right hand side are symmetric under the interchange of the $%
\theta ^{\prime }s.$ In terms of the uniformization variable $\theta _{12}$
in $F$ the transition from $in\rightarrow out$ means a change of sign via
analytic continuation i.e. without changing the charge quantum numbers $%
\alpha $ i.e. the position of the tensor factors. After accomplishing this
last step by the bose commutation relation the negative $\theta _{12}$
formfactor $F(\theta _{21})$ can according to the definition \ref{def} be
identified with $f_{21}(\theta _{2},\theta _{1})$ and the relation 
\begin{equation}
\left\langle 0\left| A(0)\right| \theta _{2}\theta _{1}\right\rangle
^{out}S(\left| \theta _{1}-\theta _{2}\right| )=\left\langle 0\left|
A(0)\right| \theta _{1}\theta _{2}\right\rangle ^{in}
\end{equation}
agrees with \ref{def} The generalization to $^{out}\left\langle \theta
_{3}...\theta _{n}\left| A(0)\right| \theta _{1}\theta _{2}\right\rangle
_{con}^{ex}$ has a problem because replacing in by out means passing from
time-ordering to anti-time-ordering but the LSZ scattering theory produces
boundary terms contributing to the connected part. Although they are absent
for theories in which the number of in-particles are conserved, it is
unclear what property of general QFT is bringing about (i) through
specialization to factorizing d=1+1 models.

On the other hand (ii) and (iii) are consequences of the following standard
crossing formula \cite{BFK} which relate the connected part of the
generalized formfactors to the analytic master function $f$: 
\begin{eqnarray}
&&_{\bar{1}}\left\langle 0\left| A(0)\right| p_{2},...,p_{n}\right\rangle
_{2...n}^{in} \\
&=&\sum_{j=2}^{n}\,_{\bar{1}}\left\langle p_{1}\mid p_{j}\right\rangle
_{j}f_{2..\hat{j}..n}^{A}S_{2j}...S_{j-1j}+f_{12...n}^{A}(\theta _{1}+i\pi
_{-}...,\theta _{n})  \nonumber \\
&=&\sum_{j=2}^{n}\,_{\bar{1}}\left\langle p_{1}\mid p_{j}\right\rangle
_{j}f_{2..\hat{j}..n}^{A}S_{jn}...S_{jj+1}+f_{2...n1}^{A}(...,\theta
_{n}\theta _{1}-i\pi _{-})  \nonumber
\end{eqnarray}
The fastes way to understand this is to draw the corresponding graphs and
remember that a positive energy particle crosses into a negative energy
antiparticle. Successive application leads to a formula which expresses the
formfactors in terms of the analytic auxiliary function f .The analytic part
of this relation gives (ii) whereas the $\delta $-function part is
responsible for (iii). A proof that these properties do not only insure
TCP-invariance (weak locality) but also Einstein causality can be given by
using JLD spectral representations \cite{SW} However the direct derivation
of the bootstrap-formfactor axioms from the principles of QFT was hitherto
not achieved. It is part of the complicated and incomplete momentum space
analyticity problem. Even the derivation of forward dispersion relations in
particle physics took several years, not to mention the derivation of the
analytic aspects \footnote{%
Only together with the (mass shell) analytic properties the crossing
symmetry aquires a physical content.} of crossing symmetry which remained
utterly incomplete. It is precisely at this point where our modular
localization approach shows its strength. To anticipate one result, it shows
that the crossing symmetry is a kind of strengthened TCP-property and that
the cyclicity it leads to is identical to the KMS-temperature ($\equiv $%
Hawking -Unruh temperature in this special case) characterization of the
(Rindler-)wedge based Hawking-Unruh effect. From our point of view the most
valuable result is that it opens for the first time the way to a new
constructive iterative (but not perturbative) approach to non-quantization
non-Lagrangean based QFT. My confidence that this may amount to more than
just another fashion rests on the observation that the tool of modular
localization comes from a refinement of TCP which, as anybody will
immediately admit, \textit{the central structure} of local QFT.

\subsection{ Modular Theory and the Formfactor Program}

In this section we will analyze the formfactor program from the viewpoint of
modular localization. To avoid complications we start with theories which
have a diagonal S-matrix. A prototype is the Ising field theory with $%
S^{(2)}=-1,S=K$ where K is the $Z_{2}$ Klein twist of the previous section.
We look first for a weakly local (not necessarily local) field in the TCP
class associated to this simple S-matrix. For this purpose we modify the
Fermi-creation and annihilation operators $a_{in}^{\#}$ of the free Majorana
field associated to the Ising field theory in order to obtain the bosonic 
\textbf{Z}$_{2}$ commutation relations. With some experience \cite{S-S} one
immediately writes the following expressions: 
\begin{eqnarray}
b(\theta ) &=&:a_{in}(\theta )e^{-i\pi \int_{-\infty }^{\theta
}a_{in}^{*}(\theta )a_{in}(\theta )d\theta }: \\
c^{*}(\theta ) &=&:a_{in}^{*}(\theta )e^{i\pi \int_{\theta }^{\infty
}a_{in}^{*}(\theta )a_{in}(\theta )d\theta }:  \nonumber
\end{eqnarray}
\begin{eqnarray}
\,\,\,B(x) &:&=\frac{1}{\sqrt{2\pi }}\int \left\{ e^{-ipx}b(\theta
)+e^{ipx}c^{*}(\theta )\right\} d\theta \quad \\
&\curvearrowright &\quad \Theta B(x)\Theta =B^{*}(-x),\quad \Theta
=S_{Ising}\cdot \Theta _{0},\,\,\,\,\Theta _{0}\equiv \Theta _{in}  \nonumber
\end{eqnarray}
This field creates wedge localized vectors in the n-particle projections of $%
\mathcal{D}(W)=H_{R}+iH_{R}$: 
\begin{equation}
B^{\#}(x_{1})....B^{\#}(x_{n})\Omega ,\quad x_{i}\in W
\end{equation}

These $b^{\#}s$ produce the $S_{Ising}$ S-matrix: 
\begin{eqnarray*}
b^{*}(\theta _{1})....b^{*}(\theta _{n})\Omega &=&S_{Ising}a^{*}(\theta
_{1})....a^{*}(\theta _{n})\Omega \\
b^{\#}(\theta ) &=&S_{Ising}a^{\#}(\theta )S_{Ising}^{*}
\end{eqnarray*}
and fulfill the Zamolodchikov algebra:

\begin{eqnarray}
b(\theta )b^{*}(\theta ^{\prime }) &=&S^{(2)-1}(\theta -\theta ^{\prime
})b^{*}(\theta ^{\prime })b(\theta )+\delta (\theta -\theta ^{\prime })
\label{rel} \\
b(\theta )b(\theta ^{\prime }) &=&S^{(2)}(\theta -\theta ^{\prime })b(\theta
^{\prime })b(\theta )  \nonumber
\end{eqnarray}

Operators of this kind have been used a long time ago in order to exemplify
the fact that massive d=1+1 theories describe ''statistical schizons'' \cite
{S-S} in distinction to conformal field theory, where the statistics (in
form of field commutation relations) are inexorably linked with the fusion
law of charges\footnote{%
For this reason the widespread use of the terminology ''bosonization'' in
conformal QFT is unfortunate , but appropriate in d=1+1 massive theories.
Whereas ''bosonization'' (or ''fermionization'') in conformal QFT means that
the current algebra admits fermionic superselection sectors, in the massive
case the \textit{same sector} allows different ''pseudo''statistical
descriptions since schizons only have intrinsic charges but not intrinsic
statistics. A periodic table of elements in a d=1+1 world does not require
fermions.}. Recently Fring \cite{Fring} and Lashkevich \cite{Lashkevich}
constructed these operators using the realization of anyionic deformations 
\cite{S-S} of Fock space d=1+1 creation- and annihilation operators $%
a^{\#}(p)$  as a guiding principle. It was through discussions I had with
Lashkevich that I later recognized that there may be an interesting relation
between the above Zamolodchikov algebra and certain auxiliary operators in
my modular localization approach.

The above expressions may be generalized to factorizable S-matrices which
are diagonal.. The corresponding b-operators are then of the form: 
\begin{equation}
b(\theta )=a_{in}(\theta )e^{i\int_{-\infty }^{\theta }\delta _{sc}(\theta
-\theta ^{\prime })a_{in}^{*}(\theta ^{\prime })a_{in}(\theta ^{\prime
})d\theta ^{\prime }}
\end{equation}
where $\delta _{sc}(\theta )$ is the scattering phase shift. Again the TCP
invariant operators and the wedge localization through the $\check{S}$%
-domain is analogous to the constant case. From those wedge localized
n-particle states one reads off the formfactors of the would be local field.
The solution is unique.

The main observations which links the bootstrap-formfactor axiomatics (i)
(ii) and (iii) with modular localization are contained in the following
statements:

\begin{itemize}
\item  The TCP-covariant (not necessarily local, but weakly local) fields $%
B(x)$ with $x=r(sinh\chi ,cosh\chi )\in W$ which generate a right wedge
algebra $\mathcal{A}(W)$ equipped with the global vacuum state (i.e. that
state which is the vacuum with respect to the global algebra $\mathcal{A}$)
result in a KMS-temperature state with the Hawking-Unruh temperature $\beta
=2\pi $. The KMS boundary condition for a correlation function of $B^{\prime
}s$ together with local fields $A(x),x\in W$: 
\begin{eqnarray}
&&\left\langle 0\left| B(r_{1},\chi _{1})B(r_{2},\chi
_{2})...A(x)...B(r_{n},\chi _{n})\right| 0\right\rangle   \label{KMS} \\
&=&\left\langle 0\left| B(r_{2},\chi _{2})..A(x)...B(r_{n},\chi
_{n})B(r_{1},\chi _{1}+2\pi i)\right| 0\right\rangle   \nonumber
\end{eqnarray}
is equivalent to the cyclicity property (ii) (the $f$ $^{A}$is obtained from
the connected part of the B-correlation function). The $2\pi $
strip-analyticity in the $\chi ^{\prime }s,$ which is provided by the
KMS-theory, translates into momentum space analyticity for the rapidity
variables $\theta _{i}$ of $f^{A}$. In particular the analyticity and the
crossing symmetry of the S-matrix is a consequence of this temperature
structure \ref{KMS}for $A=\mathbf{1}$ e.g.$\mathbf{:}$%
\begin{eqnarray}
&&\left\langle 0\left| B(r_{1},\chi _{1}+i\pi )B(r_{2},\chi
_{2})B^{*}(r_{3},\chi _{3})B^{*}(r_{4},\chi _{4})\right| 0\right\rangle  \\
&=&\left\langle 0\left| B(r_{2},\chi _{2})B^{*}(r_{3},\chi
_{3})B^{*}(r_{4},\chi _{4})B(r_{1},\chi _{1}-i\pi )\right| 0\right\rangle  
\nonumber \\
&\Leftrightarrow &\,\,S(\theta )=S(i\pi -\theta )  \nonumber
\end{eqnarray}

\item  The (improper) state vectors $B(r_{1},\chi _{1})....B(r_{n},\chi
_{n})\Omega $ are boundary values of ''half-strip'' analytic vectors: 
\[
B(r_{1},z_{1})....B(r_{n},z_{n})\Omega ,\quad \pi >Imz_{n}>....Imz_{1}>0
\]
Upon taking the n-particle component this property translates into the half
strip analyticity of the vectors: 
\[
b^{*}(z_{1})....b^{*}(z_{n})\Omega ,\quad z_{i}=\theta _{i}+i\vartheta _{i}
\]
As a result of the Zamolodchikov commutation relations of the $b^{\#\prime }s
$, we find that there is one half strip-analytic ''master''state vector $%
\psi (z_{1},....z_{n}),$ whose different boundary values correspond to
different operator orderings: 
\begin{equation}
\lim_{\stackunder{Imz_{i}\rightarrow 0}{\pi >Imz_{P(1)}>...>Imz_{P(n)}>0}%
}\psi (z_{1}...z_{n})=b^{*}(\theta _{P(1)})...b^{*}(\theta _{P(n)})\Omega 
\end{equation}
The analytic structure in rapidity space has an intriguing similarity with
the analytic x-space structure known from Wightman's formulation of QFT. The
reason is of course the close connection between the L-boost variable $\chi $
and the rapidity $\theta .$

\item  The above state vectors generate the n-particle component of the
modular localization subspace: 
\begin{equation}
P_{n}\check{S}B(x_{1})...B(x_{n})\Omega
=P_{n}B^{*}(x_{n})...B^{*}(x_{1})\Omega 
\end{equation}
\begin{equation}
\Leftrightarrow \check{S}c^{*}(\theta _{1})...c^{*}(\theta _{n})\Omega
=b^{*}(\theta _{n})...b^{*}(\theta _{1})\Omega 
\end{equation}
\begin{equation}
\curvearrowright c^{*}(\theta _{1})...c^{*}(\theta _{n})\Omega +b^{*}(\theta
_{n})...b^{*}(\theta _{1})\Omega \in H_{R}^{(n)}(W)
\end{equation}
Here we used the relation \ref{rel} i.e. the factorizability of the theory.
As in the free case, the closed real subspace $H_{R}(W)$ represents the
encoding of the complex dense modular localization space associated with $W.$
Note that this last discussion used the factorization structure.
\end{itemize}

The generalization to charged particles and to halfinteger Lorentz spin is
straightforward but the case of nondiagonal S-matrices gives rise to
additional problems. The modular localization equation for the real subspace 
$H_{R}^{(n)}(W)$ is now: 
\begin{eqnarray}
&&\check{S}\int \psi _{\alpha _{1}...\alpha _{n}}(\theta _{1},...\theta
_{n})\left| \theta _{1},...\theta _{n}\right\rangle _{in}^{\alpha
_{1}...\alpha _{n}} \\
&=&\int (\bar{\psi}_{\alpha _{1}^{\prime }...\alpha _{n}^{\prime }}S_{\alpha
_{1}...\alpha _{n}}^{\alpha _{1}^{\prime }...\alpha _{n}^{\prime }})(\theta
_{1}+i\pi ...\theta _{n}+i\pi )\left| \theta _{1},...\theta
_{n}\right\rangle _{in}^{a_{1}...\alpha _{n}}  \nonumber \\
&=&\int \psi _{\alpha _{1}...\alpha _{n}}(\theta _{1},...\theta _{n})\left|
\theta _{1},...\theta _{n}\right\rangle _{in}^{\alpha _{1}...\alpha _{n}} 
\nonumber
\end{eqnarray}
\begin{equation}
\curvearrowright (\bar{\psi}_{\alpha _{1}^{\prime }...\alpha _{n}^{\prime
}}S_{\alpha _{1}...\alpha _{n}}^{\alpha _{1}^{\prime }...\alpha _{n}^{\prime
}})(\theta _{1}+i\pi ...\theta _{n}+i\pi )=\psi _{\alpha _{1}...\alpha
_{n}}(\theta _{1},...\theta _{n})  \label{local}
\end{equation}
This ''S-reality'' equation seems to be a new mathematical structure as far
as the mathematical physics literature is concerned. In the formulation of
crossing symmetry the charge multiplicity indices $\alpha $ must be replaced
by their conjugate values: 
\begin{equation}
S_{\alpha _{1}\alpha _{2}}^{\beta _{1}\beta _{2}}(\theta _{1}-\theta
_{2})=S_{\alpha _{2}\beta _{2}}^{\bar{\alpha}_{1}\beta _{2}}(\theta
_{2}-(\theta _{1}-i\pi ))
\end{equation}
In the case of nondiagonal $S$ one does not have operators $B(x)$ at ones
disposal. It turns out that analogous to the Bethe Ansatz inspired solution
of this problem \cite{BFK}in the formfactor approach, the modular wedge
localization (i.e. the S-reality) entails a natural Bethe Ansatz structure
which permits an explicit description of the space $H_{R}^{(n)}(W)$. The
latter is indispensable for a realization of the Zamolodchikov algebra in
the incoming Fock space and the construction of local fields within the
formfactor program. These matters will be taken up in a separate paper \cite
{SW}. There we will also present the new rich structures which result from
intersections of wedge spaces in order to describe the localization spaces
e.g. of the compact double cone regions. Although the modular structure of
wedge localized von Neumann algebras in QFT as we used it here may be found
in the literature, we do not expect most of our readers to know it.
Therefore we decided to present the modular material in a broader context 
\ref{SW}.

The coordinate-free point of view of the net theory suggests that the
formfactor program may not be the most efficient and natural way to relate
an S-matrix with local fields i.e. the inverse problem of the net field
theory: given the modular invariant $S$ of the net, reconstruct the net.
Since an S-matrix is not associated to a particular field but rather is an
invariant of a local equivalence class or a net, a direct construction (in
the spirit of the functorial construction for free systems in the
introduction) of the net instead of individual fields may be simpler than
the rather cumbersome formfactor program. Such an algebraic approach would
then consist of two parts, the conceptual problems related to the modular
wedge localization of state vectors in Fock space i.e. TCP, antiparticles,
crossing symmetry etc. and the ascend to local nets of subspaces and
associated nets of von Neumann algebras.

Apart from the new analytic modular structure, the wedge localization
equation is reminiscent of Yang's use of the Bethe Ansatz idea \cite{Yang}.
Our modular localization method therefore suggests that a suitably
generalized Bethe idea may be a valuable tool in a new constructive approach
to algebraic QFT (and not just for factorizable models). Related to this is
the hope that our approach may lead to an explicit Fock space representation
of the Zamolodchikov algebra (beyond the above diagonal cases) also in
nondiagonal factorizable models, and that one meets analogues of this
algebraic structure (which is somewhere intermediate between the algebras
generated by the Heisenberg fields and that of the incoming fields and hence
may be viewed as an algebraic QFT counterpart of the fictitious \cite{Haag}
interaction picture) in general local quantum physics.

\section{Open Ends and Outlook}

The really difficult problem in a constructive approach build on modular
localization is the passing from the net of localized subspaces to a net of
von Neumann subalgebras (educated guess: subfactors of von Neumann type $%
III_{1}$ as in the free field case). Here the most important issue is
uniqueness. The net of localized subspaces is uniquely determined by the
TCP-operator. Therefore the question of uniqueness of nets of operator
algebras can be rephrased as: does the weak locality equivalence class
contain maximally only one local Borchers class? We think that by a judicial
use of spacelike (anti)commutativity, a given S-matrix ($\Leftrightarrow $a
given $\Theta $) will maximally allow (up to isomorphisms) one field system
or algebraic net. The only argument we have in the moment is : $S=1$ $%
\curvearrowright $free field Borchers class \cite{SW}. The essential step in
the argument is the use of crossing symmetry and the Watson relations which
link physical cuts in the analytically continued formfactors (matrix
elements of a representative local field in the basis of the incoming
multiparticle spaces) to the S-matrix. The fact that analytic on shell
p-space properties, as one needs them for such arguments (e.g. crossing
symmetry as an analyticity statement and not just a suggestive formal
statement abstracted from LSZ extrapolation formulas), are difficult to
obtain from the locality and spectral principles of QFT one can try to
counteract by (as in the d=1+1 factorizable case) assuming ''maximal
analyticity'' i. e. only taking physically motivated singularities into
account. On the one hand this is certainly reasonable in a constructive
approach and on the other hand we expect that the same modular theory which
underlies the modular localization approach will also lead to a much better
understanding of on shell analytic p-space properties. The present analytic
techniques result from the so-called JLD-representation for matrix elements
of causal commutators. We expect the exploration of modular localization
concepts to give more powerful analytic results.

Some speculative remarks on the problem of associating a net of von Neumann
algebras with a net of localization spaces may be helpful at this point. In
the theory of operator algebras the Araki-Connes concept of the ''natural
cone'' allows to construct a von Neumann algebra from the knowledge of the
split of $H_{R}$ into positive cones $\pm C_{+}$: 
\begin{equation}
H_{R}=C_{+}-C_{+}
\end{equation}
In the case of factorizable models the modular localization principle which
leads to $H_{R}(W)$- and $H_{R}$(double cone)-subspaces uses the \textit{%
real (on shell) particle conservation} and gave rise to (albeit new and
subtle) quantum mechanical on shell Bethe Ansatz problems. However the local
fields (or nets of von Neumann algebras) in such theories are known to have
a very rich (non-quantum mechanical!) \textit{virtual (''off-shell'')
particle creation and annihilation structure}. The same is expected in a
theory of ''free'' anyons and plektons as opposed to free Bosons and
Fermions. According to the arguments presented in this paper, this richness
must occur in this last (unfortunately poorly understood) step from modular
localized subspaces to (Einstein-) local algebras\footnote{%
This is also step from which I would expect a profound understanding of on
shell crossing symmetry.}. Connected with this is the already mentioned
uniqueness problem i.e. the inverse scattering problem for local nets, which
we will investigate in a future publication \cite{SW}.

Finally we want to make some speculative remarks on how one imagines an
iterative approach which unlike perturbation theory (which uses interaction
densities expressed in the form of Wick-polynomials and their time-ordering)
is based on modular concepts. One would start with a Heisenberg Ansatz for a
relativistic S-matrix: 
\begin{equation}
S^{(0)}=e^{i\eta },\quad \eta =\sum \int \eta _{n}(x_{1}...x_{2}):\phi
_{in}(x_{1})...\phi _{in}(x_{n}):  \label{Hei}
\end{equation}
where the coefficient function should be Poincar\'{e}-invariant and
cluster-connected, but yet without those complicated analytic p-space
properties of multidimensional dispersion theory. In order to make at least
some formal intuitive contact with the standard perturbative implementation
of interaction via invariant free field polynomials one may choose for $\eta 
$ the same expression. This Ansatz already assures (via $\Theta $ and $J$)
the modular wedge localization ( at almost no analytic costs) which leads to
mass shell analyticity in certain rapidity variables (which form families
corresponding to the family of wedges) as in the factorizable case \ref
{reality}. However the nontriviality of intersections of wedge spaces and
the required existence of a map from the net of modular localized subspaces
to a net of von Neumann algebras suggests to iteratively correct $S^{(0)}$
by $S^{(1)},S^{(2)}..$etc. in order to achieve an \textit{increasing amount
of localization}\footnote{%
The problem seems to be vagely reminiscent of a selfconsistent Hartree-Fock
iteration with the iteratively improved interaction being the analog of the $%
S^{(n)\prime }s$, and the zero order bilinear (mean field) interaction
corresponding to an $S^{(0)}$ in the form of a Heisenberg ansatz\ref{Hei}.}.
Unlike the bootstrap-formfactor problem of the previous section for which a
candidate for a ''local'', crossing symmetric S-matrix fulfilling the
''maximal'' p-space analyticity was known at the start of the formfactor
program, such a contemplated iterative approach based on modular
localization would constitute a true (''heterotic field theory-S-matrix'') 
\textit{mixed bootstrap approach}. It would be at least as removed from
perturbation theory and functional integrals as the bootstrap-formfactor
program is from Lagrangian QFT. The interaction could then receive its name
not from Lagrangians but rather from the e.g. polynomial Ansatz for $\eta $.

At this point the question arises what, if any, is the relation between this
modular approach and the standard one build on quantization. Here it is
helpful to think of a kind of field theory-adapted \textit{''Murphy's law''}%
: if there is neither a proof nor a an intrinsic structural reason for a
conjecture in QFT, then it is wrong. Although this law does not follow from
the physical principles, I do not know a single exception to it in QFT. So
the idea, that behind the renormalized Gell-Mann Low formula or functional
integrals (even with all their instanton corrections!) or the
Bogoliubov-Shirkov axiomatics in terms of a space-time localized formal
expression $S(g),$ there is an interacting theory based on those concepts
would be an illusion, except for those low-dimensional models ($\varphi
_{2}^{4}$ etc. where there is a connection through Borel resummability) for
which one has a proof. Perturbation theory remains an infinitesimal
deformation theory with no possibility for globalization. Despite its
non-existence (invoking Murphy's law) of $S(g),$ from a formal physical
point of view it is close to the modular invariant $S$ (section 4 of \cite
{S2}. Whereas standard perturbation theory maintains the linear structures
as locality in every order but links unitarity with the non-perturbative
existence (futile according to Murphy's law in QFT), any inductive approach
based on modular properties is unitary in each step of the induction, but
acquires sharp locality only in the limit (which is also the limit in which.
the modular $S$ acquires the scattering interpretation).\newline

The present modular framework is not applicable to zero mass theories for
which the LSZ incoming fields vanish i.e. a scattering matrix cannot be
defined (''Infraparticles'' as opposed to Wigner particles). Whereas for
e.g. chiral conformal theories this is no problem since they are
scale-invariant limits of massive theories, theories as QED (which
presumably do not allow such a limiting description) require conceptual
modifications. The modular approach is uncompromising but thankful for new
conceptional challenges. It flourishes on the weaknesses, paradoxes and
contradictions of the standard approach. Even at the risk of sounding
immodest, one may hope to overcome the narrowing of QFT at the beginning of
the 70$^{ies},$ when (as a result of enforcing classically based concepts as
the ''gauge principle'' via quantization) very unfortunately noncommutative
real time field theory was forced to become a footnote of euclidean field
theory. Besides recent (in my opinion artificial) attempts to get out of the
euclidean malaise by compensating the loss with a doses of noncommutative
geometry\footnote{%
Instead of trying to escape the euclidean trap by physical brute force
methods (employing ideas from nomcommutative geometry) one should remember
that QFT has a natural noncommutative structure and \textit{allows for a
euclidean formulation and a Feynman-Kac representation only under very
special circumstances.}}, there has been a renewed interest in real time
QFT. In addition to the formfactor-bootstrap program presented in this work,
this is illustrated by recent progress on QFT in curved space-time for which
the euclidean approach is physically senseless \cite{B-F} (albeit
mathematically interesting). We hope that our ideas on modular localization
may also contribute to find a way out of the present stalemate and
particularization of QFT. In any case our confidence at the moment is more
based on its unifying point of view and its fundamental modular structure
leading to new mathematical equations with the promise of a deep relation to
Bethe Ansatz structures. Anybody with some knowledge of nonperturbative QFT
will have no problem in recognizing that one is dealing here with the most
fundamental structures which QFT has to offer.

Acknowledgments: I am indebted to Andreas Fring and Michael Karowski for
their patience in explaining some intricate points in the formfactor program
and for letting me know their explicite multi-formfactor constructions for
certain models with non-diagonal S-matrix prior to publication. I also thank
Hans-Werner Wiesbrock for his active interest in these constructive
applications of modular ideas as well as Karl-Henning Rehren for a critical
reading of a preliminary version of the manuscript.

\end{document}